\documentclass[10pt,conference,letterpaper]{IEEEtran}

\usepackage[utf8]{inputenc}
\usepackage[T1]{fontenc}
\usepackage[american]{babel}
\usepackage{microtype}
\usepackage{cite, url}
\usepackage{amsmath}
\usepackage{amsfonts}
\usepackage{amssymb}
\usepackage{eurosym}
\usepackage{bm}
\usepackage{paralist}
\usepackage{graphicx}
\usepackage{float}
\usepackage{stfloats}
\graphicspath{{figures/}}
\usepackage{tikz}
\usetikzlibrary{arrows.meta,positioning,decorations.pathreplacing,calc}

\usepackage[caption=false, font=footnotesize]{subfig}
\usepackage[per-mode=symbol]{siunitx}
\usepackage{booktabs}
\usepackage{pifont}

\usepackage{textcomp}
\usepackage{xspace}
\usepackage[capitalise]{cleveref}
\crefname{section}{Sect.}{Sects.}
\crefname{figure}{Fig.}{Figs.}
\crefname{table}{Tab.}{Tabs.}
\crefname{equation}{Eq.}{Eqs.}
\Crefname{section}{Sect.}{Sects.}
\Crefname{figure}{Fig.}{Figs.}
\Crefname{table}{Tab.}{Tabs.}
\Crefname{equation}{Eq.}{Eqs.}
\def\figname{\csname cref@figure@name\endcsname\xspace}
\def\tabname{\csname cref@table@name\endcsname\xspace}
\def\secname{\csname cref@section@name\endcsname\xspace}
\def\eqname{\csname cref@equation@name\endcsname\xspace}
\def\eqpname{\csname cref@equation@name@plural\endcsname\xspace}
\usepackage{multirow}
\newcommand{\rotext}[1]{\begin{tabular}{@{}c@{}}#1\end{tabular}}
\usepackage[ragged,flushmargin]{footmisc}
\usepackage{csquotes}

\usepackage[textsize=scriptsize]{todonotes}
\presetkeys{todonotes}{inline}{}

\usepackage{ftnright}

\usepackage{color}

\DeclareSIUnit \belm{Bm}
\DeclareSIUnit \beli{Bi}

\RequirePackage{xstring}
\RequirePackage{xparse}
\RequirePackage{acro}   
\NewDocumentCommand\acrodef{mO{#1}mG{}}{\DeclareAcronym{#1}{short={#2}, long={#3}, #4}}

\acrodef{3GPP}{3rd Generation Partnership Project}
\acrodef{4G}{4-th Generation}
\acrodef{5G}{5-th Generation}
\acrodef{6G}{6-th Generation}
\acrodef{5GV2X}[5G-V2X]{5-th Generation Cellular V2X}
\acrodef{ACC}{Adaptive Cruise Control}{long-plural=lers}
\acrodef{ADAS}{Advanced Driver Assistance Systems}
\acrodef{API}{Application Programming Interface}
\acrodef{BEV}{Battery Electric Vehicle}
\acrodef{BSM}{Basic Safety Message}
\acrodef{CACC}{Cooperative Adaptive Cruise Control}{long-plural=lers}
\acrodef{CAM}{Cooperative Awareness Message}
\acrodef{CAV}{Cooperative Automated Vehicle}
\acrodef{CV2X}[C-V2X]{Cellular V2X}
\acrodef{D2D}{Device-to-Device}
\acrodef{DENM}{Decentralized Environmental Message}
\acrodef{DIC}{Dynamic Inductive Charging}
\acrodef{DSRC}{Direct Short Range Communications}
\acrodef{FCC}{Federal Communications Commission}
\acrodef{FOT}{Field Operational Test} 
\acrodef{FSM}{Finite State Machine}
\acrodef{IDM}{Intelligent Driver Model}
\acrodef{ITS}{Intelligent Transportation System}
\acrodef{IVC}{Inter Vehicle Communication}
\acrodef{LED}{Light Emitting Diode}
\acrodef{LTE}{Long Term Evolution}
\acrodef{LTEV2X}[LTE-V2X]{Cellular V2X}
\acrodef{MAC}{Medium Access Control}
\acrodef{MEC}{Mobile Edge Computing}
\acrodef{MPC}{Model Predictive Control}{short-indefinite={an}}
\acrodef{PDR}{Packet Delivery Ratio}
\acrodef{QP}{Quadratic Program}
\acrodef{RSU}{Road Side Unit}
\acrodef{SAE}{Society of Automotive Engineering}
\acrodef{SINR}{Signal to Interference plus Noise Ratio}{short-indefinite={an}}
\acrodef{SIR}{Signal to Interference Ratio}
\acrodef{SNR}{Signal to Noise Ratio}
\acrodef{SoC}{State-of-Charge}
\acrodef{SUMO}{\textsl{Simulation of Urban MObility}}
\acrodef{TA}{Traffic Authority}
\acrodef{TraCI}{Traffic Control Interface}
\acrodef{UE}{User Equipment}
\acrodef{URLLC}{Ultra-Reliable Low-Latency Communication}  
\acrodef{Uu}{User to Network Interface}
\acrodef{V2V}{Vehicle to Vehicle communication}
\acrodef{V2X}{Vehicle to Everything}
\acrodef{VANET}{Vehicular Ad-hoc Network}
\acrodef{VLC}{Visible Light Communication}
\acrodef{VUT}{\textsl{Vehicle Under Test}}
\acrodef{VRU}{Vulnerable Road Users}
\acrodef{ZOH}{Zero Order Hold}
\acrodef{VPM}{Vehicle per Minute}

\newenvironment{projectacks}
  {\par\scriptsize}
  {\par\addvspace{\bigskipamount}}

\DeclareSIUnit \km {\kilo\meter}
\DeclareSIUnit \kmh {\kilo\meter\per\hour}
\DeclareSIUnit \mps {\meter\per\second}
\DeclareSIUnit \mpsq {\meter\per\second\squared}
\DeclareSIUnit \mpsc {\meter\per\cubic\second}

\def\BibTeX{{\rm B\kern-.05em{\sc i\kern-.025em b}\kern-.08em
    T\kern-.1667em\lower.7ex\hbox{E}\kern-.125emX}}

\newcommand{\itsp}{ITS-PM\xspace}

\makeatletter
\def\ps@IEEEtitlepagestyle{%
  \def\@oddfoot{\copyrightnotice}%
  \def\@evenfoot{}%
}
\makeatother

\newcommand{\copyrightnotice}{%
  \begin{minipage}{\textwidth}
  \footnotesize \copyright\ 2026 IEEE. Personal use of this material is permitted. Permission from IEEE must be obtained for all other uses, in any current or future media, including reprinting/republishing this material for advertising or promotional purposes, creating new collective works, for resale or redistribution to servers or lists, or reuse of any copyrighted component of this work in other works.
  \end{minipage}
}

\newcommand\DeltaT{\ensuremath{\Delta T}\xspace}

\newcommand\tauv{\ensuremath{\tau_v}\xspace}
\newcommand\taumin{\ensuremath{\tau_{\min}}\xspace}
\newcommand\Pst{\ensuremath{P_{s,t}}}
\newcommand\Vt{\ensuremath{\mathcal{V}_t}\xspace} 
\newcommand\Sset{\ensuremath{\mathcal{S}}\xspace} 

\newcommand\Pvt{\ensuremath{P_{v,t}}\xspace}      
\newcommand\sit{\ensuremath{b_{v,t}}\xspace}      

\newcommand\Ei{\ensuremath{B_v^{\max}}\xspace}           
\newcommand\poni{\ensuremath{P_v^{\mathrm{pad}}}\xspace}  
\newcommand\socvhorzero{\ensuremath{b_{v,t}}\xspace}   
\newcommand\socvinit{\ensuremath{b_{v,t_{in}}}\xspace}   

\newcommand{\Pdrv}{\ensuremath{P_{\mathrm{drv}}}\xspace}  

\newcommand\sidesi{\ensuremath{b_v^{\mathrm{des}}}\xspace} 

\newcommand\siExi{\ensuremath{b_{v,t_{out}}}\xspace}   
\newcommand\Cit{\ensuremath{C_{v,t}}\xspace}      
\newcommand{\vehocc}{\ensuremath{v_o}\xspace}      

\newcommand{\stripe}{\ensuremath{s}\xspace}  
\newcommand{\stripeset}{\ensuremath{\mathcal{S}}\xspace}  
\newcommand{\stripel}{\ensuremath{L_s}\xspace}  
\newcommand{\eff}{\ensuremath{\eta}\xspace}    
\newcommand{\lam}{\ensuremath{\zeta}\xspace}    
\newcommand{\PcoilNom}{\ensuremath{P_{\mathrm{coil}}}\xspace}  
\newcommand{\TotPows}{\ensuremath{P_{\mathrm{tot},s}}\xspace}     
\newcommand{\DelPowc}{\ensuremath{P_{\max,s}}\xspace}     
\newcommand{\TotPow}{\ensuremath{P_{\mathrm{tot}}}\xspace}         
\newcommand{\numstripes}{\ensuremath{|\stripeset|}} 
\newcommand{\coilsize}{\ensuremath{d_c}\xspace}      

\newcommand\gammasvt{\ensuremath{\ensuremath{\gamma_{v,s,t}}}\xspace} 
\newcommand\gammasvk{\ensuremath{\ensuremath{\gamma_{v,s,k}}}\xspace} 

\newcommand\ui{\ensuremath{u_{v,t}}\xspace}

\newcommand\socfullfil{\ensuremath{\phi_v}\xspace} 

\IEEEoverridecommandlockouts

\begin{document}

\title{A MEC-Based Optimization Framework\\ for Dynamic Inductive Charging}

\author{
    \IEEEauthorblockN{Emre Akıskalıo\u{g}lu\IEEEauthorrefmark{1}, Mustafa Atmaca\IEEEauthorrefmark{1}, Lorenzo Ghiro\IEEEauthorrefmark{2}\IEEEauthorrefmark{3}, Giovanni Perin\IEEEauthorrefmark{2}\IEEEauthorrefmark{3}, and Renato Lo Cigno\IEEEauthorrefmark{2}\IEEEauthorrefmark{3}}
    \IEEEauthorblockA{\IEEEauthorrefmark{1}Department of Mechanical Engineering, Faculty of Technology, Marmara University, Maltepe/İstanbul, Turkey}
    \IEEEauthorblockA{\IEEEauthorrefmark{2}Department of Information Engineering, University of Brescia, Brescia, Italy}
    \IEEEauthorblockA{\IEEEauthorrefmark{3}National Inter-University Consortium for Telecommunications (CNIT), Parma, Italy}
}

\maketitle


\begin{abstract}
Range anxiety and long recharging times remain critical barriers to electric vehicle adoption. 
\ac{DIC} offers a compelling solution by enabling wireless power transfer while driving, potentially reducing battery size requirements and thus vehicle costs. 
However, \ac{DIC} infrastructures are expensive and power-constrained, requiring intelligent resource allocation to maximize user satisfaction and economic viability.
We propose a Model Predictive Control framework for optimal power allocation in \ac{DIC} systems, using edge computing and vehicular communications to prioritize vehicles with critical battery states. 
The framework is implemented and evaluated through SUMO-based simulations on a realistic 10\,km urban scenario in Istanbul, Turkey, under varying traffic intensities.
Results demonstrate two critical limitations of uncoordinated allocation.
First, resource utilization remains suboptimal despite available power when demand saturates system capacity.
Second, when demand exceeds capacity, uniform distribution of power leaves a heavy tail of critically unsatisfied vehicles that may require emergency stops.
Our MPC-based strategy addresses both regimes---maximizing power utilization during saturation through dynamic stripe rebalancing, 
and improving satisfaction fairness under scarcity by aggressively prioritizing depleted batteries at the expense of well-charged vehicles.
The framework and simulation tools are released as open-source to support further research in this emerging domain.
\end{abstract}


\acresetall

%

\section{Introduction}
\label{sec:introduction}

Range anxiety remains a critical barrier to widespread \ac{BEV} adoption.
While stationary charging infrastructure is rapidly expanding, the persistent need for lengthy stops and strategic route planning continues to limit the appeal of electric mobility.
\ac{DIC}~\cite{sagar23-access, farghly25-access, bouanou2025PowerSources} offers a paradigm shift by enabling vehicles to charge wirelessly while driving, eliminating range constraints and potentially reducing battery size requirements and costs~\cite{yeong2015-economic}.

The technology is transitioning from laboratory experiments to real-world deployment.
Field operational tests are demonstrating viability~\cite{electreon_arena_of_the_future, mdot_wireless_charging_roadway}, with the KAIST OLEV project~\cite{olev2015} operating continuously since 2011 and expanding beyond campus boundaries.
Although infrastructure costs substantially exceed those of conventional charging stations, high utilization on trafficked roads can justify the investment.
The technology itself continues to evolve rapidly, with ongoing research addressing efficiency, coil alignment robustness, and grid integration~\cite{sagar23-access, farghly25-access, bouanou2025PowerSources}.

However, a critical gap exists in the literature: While extensive research optimizes single-vehicle charging performance---essential for technical viability---
the system-level behavior under realistic multi-vehicle traffic remains unexplored.
\ac{DIC} infrastructures are fundamentally power-constrained shared resources where demand can significantly exceed capacity during peak periods.
This raises a central question: how should limited power be allocated across vehicles with heterogeneous charging needs to improve both system efficiency and user satisfaction?

This work addresses this gap by investigating, for the first time to the best of our knowledge, how \ac{MEC}-based optimal power allocation can maximize both user satisfaction and operator revenues in realistic urban scenarios.
We develop a \ac{MPC} framework that dynamically prioritizes vehicles with critical battery states learned through V2X communications,
and evaluate its performance using a realistic \SI{10}{\km} urban network in Istanbul, Turkey, modeled in SUMO\footnote{\ac{SUMO} is a popular open-source traffic simulator; Homepage: \url{https://eclipse.dev/sumo}}.

This paper makes three key contributions:
\begin{itemize}
\item An \ac{MPC}-based power allocation algorithm that maximizes satisfaction fairness under resource scarcity by dynamically balancing individual vehicle urgency, predicted trajectories, and system-wide capacity constraints;
\item A communication protocol design for information exchange between infrastructure and vehicles to support the optimization framework;
\item An open-source SUMO-based simulation framework that captures realistic \ac{DIC} constraints, including stripe topology, power budgets, and traffic variability, enabling reproducible research in this emerging domain.
\end{itemize}

%

\section{Background and Motivation}
\label{sec:related}

\acp{BEV} are expected to play a dominant role, and vehicles are increasingly connected: \acs{5G} and \acs{6G} networks position \ac{V2X} as one of the centerpieces of the architecture. 
Here, we provide a brief overview of recent advances in wireless inductive charging and discuss the role of vehicular communications, the two pillars supporting this~work. 

\subsection{Dynamic Inductive Charging}
\label{ss:dic}

\ac{DIC} allows \acp{BEV} to receive energy while driving from inductive coils embedded beneath the road to vehicle-mounted pads. 
Recent studies indicate that \ac{DIC} can complement conventional charging by mitigating range anxiety and reducing the need for large on-board batteries, particularly in dense urban environments \cite{Nguyen2024DIC}.
Unlike static charging stations, \ac{DIC} infrastructures operate as shared and power-limited systems. 
Individual inductive coils are grouped into discrete \textit{stripes}---road segments where strings of coils are deployed contiguously and share the power supply over limited distances.
Continuous installation along entire roads is impractical: road geometry, intersection topology, and structural complexity impose physical constraints on where coils can be embedded, while high installation costs favor strategic deployment in sections with sufficient traffic density \cite{Hwang2018SystemOpt}.
Due to this strong coupling between mobility and energy transfer, the efficiency of \ac{DIC} infrastructures cannot be evaluated independently of traffic conditions. 
Researchers rely on microscopic traffic simulators to represent vehicle-level dynamics and energy interactions within urban mobility contexts, which include the examination of charging usage, equity, and congestion impacts on a larger scale~\cite{Koch2021SUMO, Ren2024EcoDriving}.
Existing simulation-based studies have primarily focused on infrastructure planning---such as optimal stripe placement under installation budget constraints \cite{apicella24-energies}---while operational strategies for power allocation given deployed infrastructure remain largely unexplored.
Uncoordinated power allocation leads to inefficient resource utilization under high traffic load, motivating the adaptive and context-aware control mechanisms we investigate in this work.

\subsection{Communication Strategies}
\label{ss:cost}

Efficient operation of \ac{DIC} requires low-latency \ac{V2X} communications to exchange charging requests and context information, as well as power allocation messages. 
Without coordination, vehicles will recharge from a \ac{DIC} stripe regardless of their needs and battery status (unless it is full or the driver prevents it). 
While this is acceptable when there is sufficient power for all vehicles, when traffic is dense and the total \ac{DIC} power is less than the power requested by vehicles, suboptimal allocation may lead to user dissatisfaction or even battery depletion for some vehicles. 
The extension of the ETSI ITS-G5 standard to support messages for \ac{DIC} optimization is simple, as existing messages can be used and combined to deliver the information required from vehicles to the infrastructure, while power assignment would require additional dedicated messages.  
As \ac{DIC} optimization does not have safety implications, nor real-time or multicast constraints, standard \ac{URLLC} is sufficient to support the application in the framework of 5G or 6G networks. 
Collecting requests, running the power allocation algorithm, and sending back results must be done in a few hundred milliseconds at most, thus a \ac{MEC} solution is preferred to cloud-based ones as discussed in \cite{Cao2018MEC}. 
Even standard wired charging suffers from poor performance when run without proper coordination, so that edge communications and computing are proposed to improve operators' revenues \cite{Zhang2021EdgeIntelligence}. 

\Cref{sec:protocol} further discusses the communication needs, protocol, and architecture required to support the proposed framework.

%

\section{Scenario}
\label{sec:scenario}

\begin{figure}
\centering
\includegraphics[width=\columnwidth]{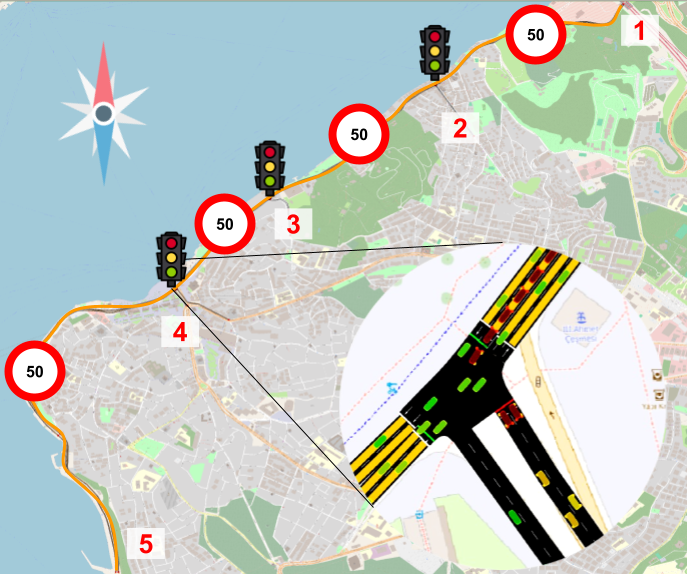} 
\caption{The Uskudar, Istanbul, road modeled for this study.
The main road is \SI{9.65}{\kilo\meter} long, with two lanes per direction.
Stripes, superimposed as an orange thick line, are deployed along this main road.
Entry/Exit points are numbered from 1 to 5 from North to South, and the traffic relations are reported in \cref{tab:trf}.
Three distinct traffic intensities are considered by choosing $\lambda \in \{5, 12, 20\}$ \acf{VPM}.
}
\label{fig:uskudar}
\vspace{-3mm}
\end{figure}

\begin{table}
\caption{\label{tab:trf}
Vehicular traffic relations ($\lambda$ [vpm]).}
\vspace{-4mm}
\begin{center} 
\begin{tabular}{lc|c|c|c|c|c|} 
 \multicolumn{2}{c}{~~~} & \multicolumn{5}{c}{Exit} \\ 
 \multicolumn{2}{c|}{~~~} &   1   &   2   &    3   &   4   &    5   \\   
\cline{2-7} 
\multirow{5}{*}{\rotext{\rotatebox[origin=c]{90}{Entry}}} 
 & 1  &   --            & $\lambda/8$ & $\lambda/8$ & $\lambda/4$ & $\lambda/2$ \\
\cline{2-7} 
 & 2 & $\lambda/8$ &    --           &    --          &    --           & $\lambda/8$ \\
\cline{2-7} 
 & 3  & $\lambda/8$ &    --           &    --          &    --           & $\lambda/8$ \\
\cline{2-7} 
 & 4 & $\lambda/4$ &    --           &    --          &    --           & $\lambda/4$ \\
\cline{2-7} 
 & 5  & $\lambda/2$ & $\lambda/8$ & $\lambda/8$ & $\lambda/4$ &  --         \\
\cline{2-7} 
\end{tabular}
\end{center}
\vspace{-6mm}
\end{table}

As a case study, we have selected an approximately $\SI{9.65}{\km}$ stretch of a slow urban road on the coast of the Bosphorus in Istanbul, Turkey. 
The road is mostly a double carriageway with two lanes per direction, and the average speed is below \SI{40}{\kmh} (speed limit \SI{50}{\kmh}), with very high intensity. 
This makes it an ideal candidate for inductive recharging: high vehicle density; limited space for standard recharging stations; and urgent need for pollution and CO$_2$ reduction in dense urban environments where \ac{BEV} adoption is critical. 

\Cref{fig:uskudar} reports the area taken from OpenStreetMap with superimposed the \ac{SUMO} road model.
The modeled network includes the main \SI{9.65}{\kilo\meter} road segment together with the major signalized junctions imported from OpenStreetMap. 
Minor intersections are omitted. 
The main road geometry was slightly regularized by widening a small number of narrow sections that generated localized bottlenecks in preliminary simulations.

All intersections are equipped with adaptive traffic lights that automatically regulate the green-yellow-red cycles to optimize throughput, as represented in the zoom of the \ac{SUMO} topology, including some passing vehicles reported in the inset of \cref{fig:uskudar}. 
Entry/Exit points are labeled \num{1} to \num{5} from North to South; \cref{tab:trf} reports the traffic relations relative intensity; the traffic intensity $\lambda$ is defined in Vehicles per Minute [vpm].
For the sake of simplicity and balance, we avoid traffic between non-terminal intersections, assigning higher traffic to larger intersections (see \cref{tab:trf}). 

Both lanes of each carriageway are equipped with realistic \ac{DIC} \textit{stripes} (denoted $s \in \Sset$) composed of small charging elements called \textit{coils}, so that $s$ is a $n_s$-long sequence of coils $s = \{c_0, c_1, \ldots, c_{n_s} \}$, as small coils are more efficient to guarantee a good inductive coupling \cite{patil2018electrification}. 
Since we are not concerned with the charging technology, we do not consider details of the coils composing the pads (circular, rectangular, compensated, etc.) nor the inefficiencies due to misalignments of the road coils with the on-board coils. 
Coils are dense, so we can assume charging is constant while the vehicle is moving or standing still at a traffic light. 
Each coil can deliver a maximum nominal power $\PcoilNom=\SI{100}{\kilo\watt}$ with an efficiency \eff we assume constant (e.g., 95\%); 
the vehicle charging capability exceeds the per-coil supply (\PcoilNom), so the system is limited by coils and stripes and not by vehicles. 
A stripe $s$ is limited by the topology of the road: Coils are difficult if at all possible to deploy in intersections where vehicle trajectories are less predictable; it is not recommended to have coils within \SI{5}{\meter} from pedestrian crossings; and so forth.

In the considered topology there are $\numstripes=10$ stripes (\num{5} per direction) with lengths ranging from \SI{628}{\meter} to \SI{1276}{\meter}.
Each stripe $s$ can deliver at most a power 
\begin{equation}
\DelPowc = 2 \cdot \left\lfloor \frac{\stripel}{\coilsize} \right\rfloor \cdot \eta \PcoilNom,
\label{eq:totps}
\end{equation}
where \stripel is the length of stripe \stripe in [m] and \coilsize, also in [m], is the distance between coils centers; the factor 2 accounts for the 2 lanes of the road (i.e., each stripe has 2 coil strings, one in each lane).
This potential delivered power is not achievable in practice, as, on a road, there will always be more coils than actively charging vehicles. 
In more formal terms, it always holds that $\coilsize < \vehocc$, where \vehocc [m] is the average longitudinal road occupancy of a vehicle, including both vehicle length and spacing. 
Since $\coilsize < \vehocc$, the space occupied by a vehicle spans multiple coils, and the vehicle can couple only with the subset of coils that lie within the active coupling region of its on-board charging pad.
Therefore, the aggregate nominal power obtained by summing the contribution of all coils along the stripe is not achievable in practice.

For modeling simplicity in this initial study we set $\coilsize=\SI{1}{\meter} $ and $\vehocc=\SI{10}{\meter}$ equal for all stripes, assuming that on a slow moving road vehicles maintain an average bumper-to-bumper distance of \SI{5}{\meter} and that vehicles are \SI{5}{\meter} long on average.

The total power available for the \ac{DIC} system is much smaller than the potential delivered power: 
\begin{equation}
\TotPow \ll \sum_{\stripe\in\stripeset} \DelPowc 
\label{eq:totpow}
\end{equation}
The power management system can freely allocate the power to each stripe respecting \cref{eq:totps,eq:totpow}. 
If no active power management is available, \TotPow is statically split between stripes based on the length \stripel of the stripes themselves, assigning $\TotPows < \DelPowc$ to each of them. 
\Cref{tab:notation} reports the main notation used in this work. 
\begin{table}[thb]
\centering
\caption{Summary of notation used in \cref{sec:scenario,sec:opt}.}
\label{tab:notation}
\begin{tabular}{ll}
\toprule
\textbf{Symbol} & \textbf{Description} \\
\midrule
\multicolumn{2}{l}{\textit{Infrastructure and Charging System}} \\
$s, \Sset$ & strip, set of \ac{DIC} stripes \\
$\PcoilNom$ & Nominal power per coil [kW] \\
$\DelPowc$ & Maximum deliverable power of stripe $s$ [kW]\\
$\TotPow$ & Total \ac{DIC} system power budget [kW] \\
$\TotPows$ & Power assigned to stripe $s$ w/o power management [kW] \\
$P_{\rm{min}, s}$ & Minimum power for stripe $s$ [kW] \\[2mm]
\multicolumn{2}{l}{\textit{Vehicles and Battery}} \\
$v, \Vt $ & Vehicle index, Set of vehicles at time $t$ \\
$\Ei$ & Battery capacity of vehicle $v$ [kWh] \\
$\poni$ & Maximum power absorbable by $v$ [kW] \\
$\sit$ & SoC of vehicle $v$ at time $t$ [\%] \\
$\socvinit$ & SoC of vehicle $v$ at \ac{DIC} area entry [\%] \\
$\sidesi$ & Desired SoC at \ac{DIC} area exit [\%] \\
$\socvhorzero$ & Current SoC at start of MPC horizon [\%] \\
$\siExi$ & Actual SoC at \ac{DIC} area exit [\%] \\
$\socfullfil$ & Satisfaction metric for vehicle $v$ [\%] \\
$\Pdrv$ & Power required for driving [kW] \\
$\Cit$ & Predicted energy consumption in time interval $t$ \\[2mm]
\multicolumn{2}{l}{\textit{Power Allocation and Dynamics}} \\
$\Pvt$ ($\Pst$) & Power allocated to vehicle $v$ (stripe $s$) at time $t$ [kW]\\
$P^{r}_{v,t}$ & Power requested by vehicle $v$ at time $t$ [kW] \\
$P^{r}_{s,t}$ & Total power requested on stripe $s$ at time $t$ [kW] \\
$\gammasvt$ & Indicator: vehicle $v$ on stripe $s$ at time $t$ \\[2mm]
\multicolumn{2}{l}{\textit{MPC Framework}} \\
$K$ & Prediction horizon length [num] \\
$\DeltaT$ & Control time interval [s] \\
$b_{v,k}$ & Predicted \acs{SoC} of vehicle $v$ at step $k$ [\%] \\
$P_{v,k}$ ($P_{s,k}$) & Power assigned to vehicle $v$ (stripe $s$) at step $k$ [kW] \\
$\mathcal{J}_t$ & MPC objective function at time $t$ \\
$\ui$ & Urgency weight for vehicle $v$ at time $t$ \\
$\tauv$ & Remaining time before vehicle $v$ exits [s] \\ 
$\lam$ & Trade-off coefficient: satisfaction vs.\ revenue \\
$\xi$ & Weight for stripe power assignment \\
$r_{v,k}$ & Energy price for vehicle $v$ at step $k$ [arbitrary units] \\
\bottomrule
\end{tabular}
\end{table}

\subsection{Total Requested Power} 
\label{ss:trp} 

If there is no active management of the recharging energy assigned to each vehicle, then each vehicle at any time tries to draw from the recharging system a power
\begin{equation}
P^{r}_{v,t} = \min\left(\poni,\PcoilNom, \left[\frac{[\sidesi - \socvhorzero]\Ei}{T_{c}}+ \Pdrv  \right] \right) \cdot \gammasvt
\label{eq:reqp}	
\end{equation}
where:
\begin{inparaenum}[(i)]
 \item \gammasvt is an indication function telling if the vehicle $v$ is on stripe $s$ at time $t$; 
 \item $T_{c}$ is a constant time modulating the recharge power when close to the desired level;
 \item $\Pdrv$ is the power required by the engine to drive; and
 \item \sidesi is the desired target \ac{SoC} and \sit the current \ac{SoC}. 
\end{inparaenum}
The total requested power on a stripe $s$ at time $t$ is 
\begin{equation}
P^{r}_{s,t} = \sum_v P^{r}_{v,t} 
\label{eq:reqps}
\end{equation}
and the total power requested to the \ac{DIC} system 
\begin{equation}
P^r_{tot,t} = \sum_s P^r_{s,t} 
\label{eq:reqpt}
\end{equation}
Importantly, the total power requested on stripe $s$ ($P^r_{s,t}$) can exceed that stripe's maximum deliverable power $\DelPowc$,
and the total system demand ($P^r_{tot,t}$) can likewise exceed the system's total available budget $\TotPow$.
These scenarios represent two critical resource scarcity regimes: local saturation on individual stripes and global saturation of the entire system.
The quantities $P^r_{s,t}$ and $P^r_{tot,t}$ thus measure demand that may not be satisfiable given the physical constraints on power delivery.  

%

\section{Optimized Power Allocation} 
\label{sec:opt}

\begin{table*}[!b]
\centering
\hrule
\begin{equation}
\label{eq:obj}
\mathcal J_t\left(\{b_{v,k}\}_{k=t}^{t+K}, \{P_{v,k}, P_{s,k}\}_{k=t}^{t+K-1}\right)
=
\underbrace{\sum_{k=t}^{t+K}\sum_{v\in\Vt}
u_{v,k}(b_{v,k}-\sidesi)^2}_{\text{desired battery tracking}}
- \lam
\underbrace{\sum_{k=t}^{t+K-1}\sum_{v\in\Vt}
r_{v,k}P_{v,k}\DeltaT}_{\text{ITS revenue}}
+ \xi
\underbrace{\sum_{k=t}^{t+K-1}\sum_{s\in\Sset}
\left(P_{s,k}- \sum_{v\in\mathcal V_t}
\gammasvk \, P_{v,k}\right)^2}_{\text{stripe power assignment}}
\tag{13}
\end{equation}
\end{table*}

Uncoordinated charging can reduce the residual range of vehicles with depleted batteries and limit potential revenues. 
We propose here a power allocation optimization based on \acf{MPC} to maximize user satisfaction and operator revenues.  
The approach consists of defining an objective function to control the \emph{state variables} of the system to a target value using \emph{control variables}. 
In our context, we control the vehicles' battery \acp{SoC} by supplying power from the \ac{DIC} infrastructure. 
The optimization problem is solved over a finite predictive \emph{time horizon}. 
Following a \emph{receding window} approach, only the first computed control is applied, then the window slides to the next discrete time interval \(\DeltaT\), repeating the procedure.
Index $k$ denotes the prediction step, and the time horizon has length $K$ steps. 
The predictive horizon is thus $K \cdot \DeltaT~\rm [s]$.

\subsection{Decision variables and constraints}
\label{ss:dvc}

We consider three sets of optimization variables:
\begin{inparaenum}[(i)]
    \item the per-vehicle charging powers $P_{v,k}$, for \(k=\{t,\dots,t+K-1\}\),
    \item the power assigned to stripes $P_{s,k}$, for \(k=\{t,\ldots,t+K-1\}\), and
    \item the predicted battery \acp{SoC} $b_{v,k}\in[0,1]$, for \(k=\{t,\dots,t+K\}\).
\end{inparaenum}
Coil availability is encoded by \(\gammasvt\in\{0,1\}\), predicting if vehicle $v$ can couple with some coil of the stripe $s$ at time $t$.
The prediction relies on a simple extrapolation from current average travel times.

\subsubsection{Battery dynamics}
Let \(\Cit\) be the predicted energy consumption in $[t, t+\DeltaT]$ for every vehicle \(v\in\Vt\).
The battery \ac{SoC} evolves as: 
\begin{equation}
\label{eq:socdyn}
    b_{v,t+1} = \sit + \eta \,\frac{\Pvt\,\DeltaT}{\Ei} - \frac{\Cit}{\Ei},
\end{equation}
with bounds given by
\begin{equation}
0 \le \sit \le \sidesi.
\end{equation}

\subsubsection{Power allocation bounds}
The power allocation variable $\Pvt$ is nonnegative and is upper-bounded by $\poni$ and $\PcoilNom$. 
This amounts to adding the linear constraint
\begin{equation}
0 \le \Pvt \le \min\left\{\poni,  \PcoilNom \sum_{s\in\Sset} \gammasvt \right\},
\label{eq:pervehpower}
\end{equation}
where the indicator function \gammasvt ensures that $\Pvt=0$ if a vehicle is not coupled with any stripe $s$ at time $t$.

\subsubsection{ITS system constraints}
Each stripe of the \ac{DIC} area has an individual deliverable power bound 
\begin{equation}
\label{eq:stripe-bound}
P_{\rm{min}, s} \le \Pst \le \rho\TotPows,
\qquad \forall\, s\in\Sset,\;\; \forall\, t,
\end{equation}
where $\rho > 1$ is a coefficient that allows power transfer among stripes ($\rho = 4$ in this work); a minimum power $P_{\rm{min}, s} > 0$ is always assigned to the stripe to ensure operation. 
The vehicles that are charging on the coils of stripe $s$ cannot absorb more than the power assigned to the stripe:
\begin{equation}
\label{eq:vehicles-on-stripe}
\sum_{v\in\Vt} \gammasvt\, \Pvt \le \Pst,
\qquad \forall\, s\in\Sset,\;\; \forall\, t.
\end{equation}
Finally, the total power budget of the \ac{DIC} area must be allocated to the stripes:
\begin{equation}
    \label{eq:totalpower}    
    \sum_{s\in\Sset} \Pst =\TotPow, \qquad \forall\, t.
\end{equation}

\subsection{MPC problem formulation}

For the cost function, let us define the urgency weight
\begin{equation}
    \ui = \frac{\big[\sidesi - \socvhorzero\big]_+}{\max(\tauv,\taumin)},
\end{equation}
where \tauv is the vehicle's remaining travel time and $\taumin > 0$ is a small safety scalar. The rationale behind this choice is that we want to prioritize charging the vehicles with a large positive battery gap from the desired level (large value at the numerator) and those that are close to exiting the system (small value at the denominator).

The \ac{MPC} objective at time \(t\) is given in Eq.~\eqref{eq:obj}, where 
$r_{v,t}$ is the time-varying and vehicle-dependent energy price; the coefficient $\lam$ is used at the \ac{ITS} to trade customer satisfaction for system revenue, whereas $\xi$ controls the weight of the stripe power assignment. 
We set $r_{v,t}$ constant, so it does not influence the presented results. 

The \ac{ITS} power allocation optimization problem at time $t$ can thus be defined as: 
\setcounter{equation}{13}
\begin{equation}
    \label{eq:optimization-problem}
    \begin{aligned}
        \min_{\{b_{v,k}, P_{v,k}, P_{s,k}\}}\quad & \mathcal J_t\left(\{b_{v,k}\}_{k=t}^{t+K}, \{P_{v,k}, P_{s,k}\}_{k=t}^{t+K-1}\right) \\
        \text{s.t.} \qquad\,\quad& \eqref{eq:socdyn}-\eqref{eq:totalpower}
    \end{aligned}
\end{equation}
which is a convex \ac{QP} and can be solved to global optimality by any commercial solver.

\section{Data Exchange}
\label{sec:protocol}

\begin{table}[tbh]
\caption{\label{tab:messages}
Minimal set of parameters exchanged between vehicles and the charging infrastructure management.}
\begin{tabular}{lp{0.6\columnwidth}p{0.12\columnwidth}}
\toprule \\
\textbf{Field}      &  \textbf{Meaning}  & \textbf{Symbol} \\
\midrule 
\multicolumn{3}{c}{\textit{vehicle to \itsp}}\\
BC      & Battery Capacity [kWh] & \Ei \\
BL      & Battery Level [percent]  & \sit \\
BTE    & Battery level Target at Exit point [\%] & $\sidesi$ \\ 
PADP  & Max Power Absorbable by the on-board pad [kW] & \poni\\
ROUTE  & A list of waypoints in the future that define the vehicle's intended route & -- \\
\midrule 
\multicolumn{3}{c}{\textit{\itsp to vehicle}} \\
PCOIL &  Maximum Power Available from the coils of the stripe the vehicle is on &  \PcoilNom \\
PASS   &  Power Assigned for the next DT [kW] & $P_{v,1}$ \\
DT      &   Time to the Next Assignment [s] & $\DeltaT$ \\
PTOL   &   Tolerance of PASS [kW]; the actual delivered power is within the interval PASS$\pm$PTOL & \\
EXIT   &   The \itsp estimated point of exit from the \ac{DIC} (based on ROUTE) & \\
TEX    &   The \itsp estimated residual time to reach EXIT & $\tauv$ \\
\bottomrule
\end{tabular}
\end{table}

The optimal power allocation requires data exchange between vehicles and the \ac{ITS} subsystem that manages the \ac{DIC}, which we refer to as the \itsp, for Power Management. 
Within the framework of ETSI ITS-G5, the protocol to implement this power management scheme can be built exploiting existing messages such as \acp{CAM}, which announce position, speed, and other basic parameters, augmented by additional messages dedicated specifically to the power management. 
\Cref{tab:messages} reports a minimal set of additional parameters that must be exchanged between vehicles and \itsp to manage the power assignment. 

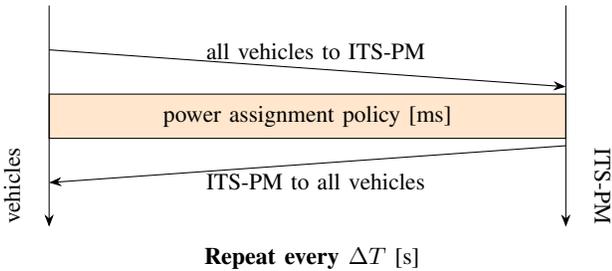
\begin{figure}[tbh]
    \centering
    \resizebox{0.95\columnwidth}{!}{%
    \begin{tikzpicture}[>=Stealth, font=\small]
        \node[rotate=90] (veh) at (0,0) [anchor=east] {vehicles};
        \node[rotate=-90] (its) at (8,0) [anchor=west] {\itsp};
        \draw[fill=orange!20,draw=black]
          ($(veh.east)+(0.5,0.6)$) rectangle ($(its.west)+(-0.5,0)$);   
        \node (opt) at ($(veh.east)!0.5!(its.west)+(0,0.3)$) {power assignment policy [ms]};
        \draw[->] ($(veh.east)+(0.5,1.8)$) -- ($(veh.east)+(0.5,-1.2)$);
        \draw[->] ($(its.west)+(-0.5,1.8)$) -- ($(its.west)+(-0.5,-1.2)$);
        \foreach \y/\i in {1.2/1} {
          \draw[->] ($(veh.east)+(0.5,\y)$) --
            node[above,pos=0.5] {\text{~~all vehicles to \itsp}}
            ($(its.west)+(-0.5,\y-0.5)$);
        }
        \foreach \y/\i in {0.4/0} {
          \draw[->] ($(its.west)+(-0.5,\y-0.5)$) --
            node[below,pos=0.5] {\text{~~\itsp to all vehicles}}
            ($(veh.east)+(0.5,\y-1)$);
        } 
    \end{tikzpicture}
    }\\
    {\small\textbf{Repeat every $\DeltaT$} [s]}
    \caption{Ideal data exchange scheme.}
    \label{fig:comm_protocol}
\end{figure}

Ideally, the protocol should implement the simple, synchronized data exchange sketched in \cref{fig:comm_protocol}, which is indeed, with some minor approximations, what has been implemented extending \ac{SUMO} as described in \cref{sec:results}. 

\begin{figure}[th]
  \centering
  \includegraphics[width=\columnwidth]{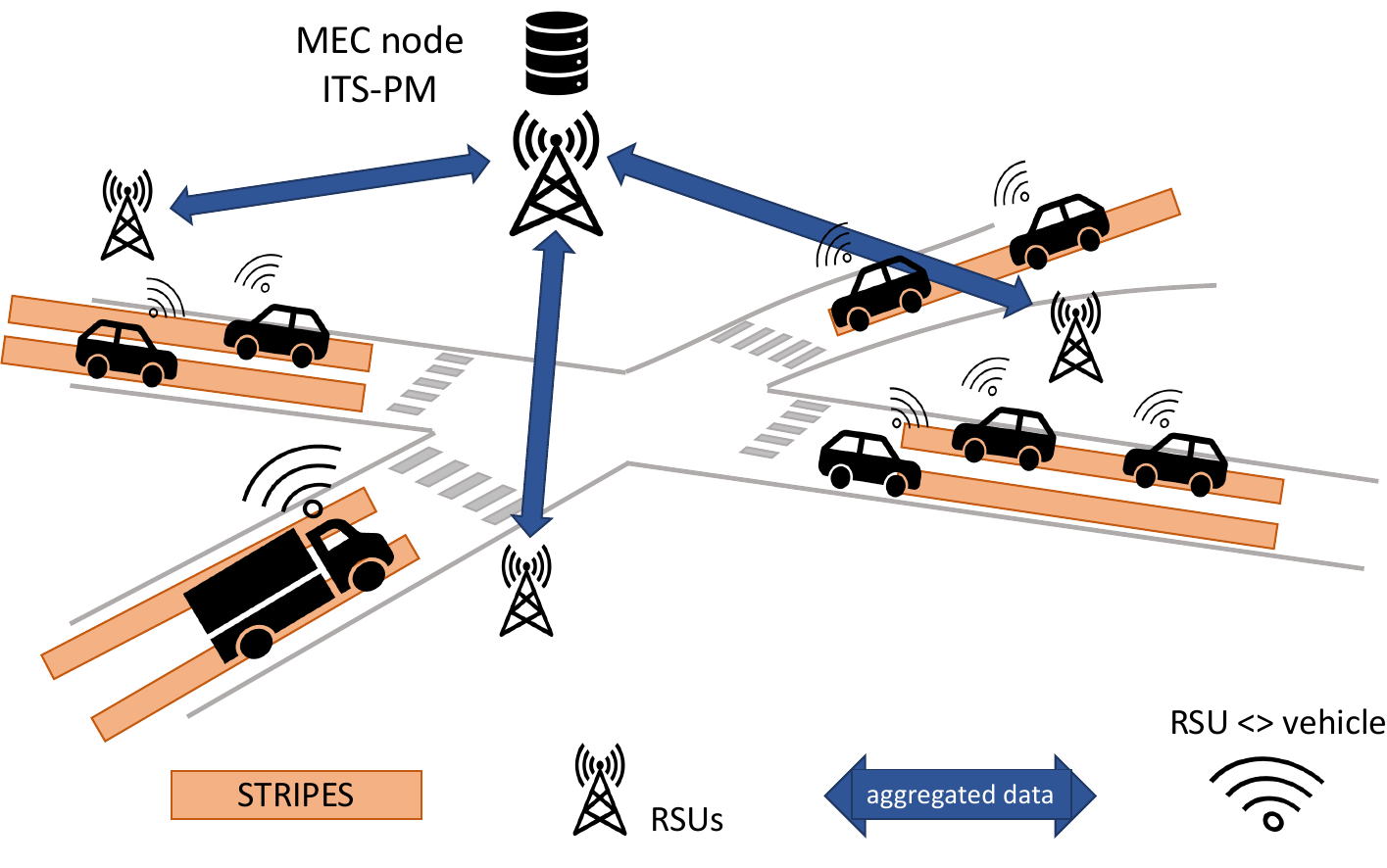}
  \caption{Realistic, \ac{MEC}-based power management system with communications.}
  \label{fig:realistic}
\end{figure}

Clearly, a realistic situation, as depicted in \cref{fig:realistic}, implies a more complex, sophisticated, and asynchronous protocol. 
The optimization procedure itself does not need to run synchronously, though it will always use, when running, a discrete-time look-ahead prediction based on a suitable $\DeltaT$. 
Similarly, information from vehicles arrives at the \itsp, implemented in a \ac{MEC} server, via multiple \acp{RSU}, possibly implemented with different technologies (802.11bd, 5G gNB, 6G, \ldots). 
\acp{CAM} are sent by vehicles following the standard generation procedure defined by ETSI, and are collected by \acp{RSU} and forwarded to \itsp, while the power management specific messages are sent on a direct logical connection directly to the \itsp, but aggregation at the \ac{RSU} is possible to spare bandwidth and other resources. 
The detailed design and performance of such a realistic system is part of our future work and goes beyond the scope of this paper. 

%

\section{Simulations and Results}
\label{sec:results}

The scenario described in \cref{sec:scenario} and the optimization framework developed in \cref{sec:opt} have been implemented as an extension of \ac{SUMO}, with Python scripts controlling the simulation advancement, emulating the communication exchange sketched in \cref{sec:protocol}, and running the optimization framework through CVXPY \cite{diamond2016-cvxpy}, which invokes Gurobi\footnote{\url{https://www.gurobi.com/lp/all/do-more-with-gurobi/}} for convex optimization.

\subsection{Simulation Setup}

We simulate three different traffic intensities, inserted in the simulation according to the traffic relations described by \cref{tab:trf}.
\begin{itemize}
\item \textbf{Low Traffic}: $\lambda=5$ [vpm] (night-time);
\item \textbf{Medium Traffic}: $\lambda=12$ [vpm] (lunch-time);
\item \textbf{Intense Traffic}: $\lambda=20$ [vpm] (rush hour around 5\,PM).
\end{itemize}

Each simulation runs for 75 minutes: a 15-minute warm-up period to converge to steady traffic conditions, followed by 60 minutes of measurement. 
Vehicles are assigned $b_{v,t_0}$ distributed in $\mathcal U(0.1, 0.5)$ and $\sidesi$ in $\mathcal U(0.5, 1.0)$, creating a realistic demand profile where the majority of vehicles require moderate to substantial charging.
The total \ac{DIC} system budget is $\TotPow=\SI{16}{\mega\watt}$, distributed across the 10 stripes with efficiency $\eta=0.95$.

To evaluate the system behavior for critically battery-depleted vehicles, we introduce specific \textsl{Vehicles Under Test} (\acsp{VUT}): vehicles that enter every 60 seconds with nearly depleted batteries and desire full recharge ($\socvinit \approx 0$ and $\sidesi=1.0$).
These vehicles represent extremely urgent charging needs with maximum battery gap, enabling us to observe how the system prioritizes critical cases; \acsp{VUT} always follow the 2 main routes, ($1 \rightarrow 5$ and $5 \rightarrow 1$).

We compare two operational strategies:
\begin{itemize}
\item \textbf{Benchmark (BENCH)}: No optimization or V2X communication.
Each vehicle charges absorbing a power limited by the \ac{DIC} constraints, as the on-board pad is over-dimensioned. 
Power is statically allocated to stripes proportionally to their length.

\item \textbf{Optimized (OPT)}: \ac{MPC}-based allocation with $\DeltaT=5$s control interval.
Vehicles communicate battery status and $\sidesi$ to the \ac{ITS}, which solves problem~\eqref{eq:optimization-problem} and communicates individual power assignments.
Moreover, dynamic power rebalancing between stripes is performed according to the MPC solution.
\end{itemize}

Overall, this results in 12 simulation configurations (12h of measurements): 3 traffic rates $\times$ 2 \ac{VUT} settings $\times$ 2 control strategies, although we can only present a few results due to space constraints.

\subsection{Performance Metrics} 
\label{ss:pm} 

A \ac{DIC} performance can be measured in many different ways. 
Global metrics include the total energy transferred to vehicles and the fraction of available power utilized, which ultimately balance revenues and investment. 
User satisfaction is equally critical, as it directly impacts service quality perception and can be leveraged to modulate energy pricing and maximize revenues through differentiated services.

Recall that $\sidesi\in[0,1]$ is the vehicle desired \ac{SoC} at exit and $\siExi\in[0,1]$ its actual \ac{SoC} at exit.
We define the \textit{\ac{SoC} Fulfillment} (\socfullfil) metric as:
\begin{equation}
\socfullfil = \frac{\siExi}{\sidesi}
\label{eq:soc_fulfillment}
\end{equation}
which measures the fraction of the desired target $\sidesi$ actually achieved at exit.
This metric captures individual vehicle satisfaction: $\phi_v = 1$ indicates complete fulfillment of charging needs, while $\phi_v < 1$ quantifies the shortfall.
Unlike absolute energy-based metrics, \ac{SoC} fulfillment is normalized to vehicle-specific targets and fairly compares satisfaction across vehicles with different battery capacities and desired charge levels.

\subsection{Framework validation}

\begin{figure*}[t]
  \centering
  \includegraphics[width=0.99\textwidth]{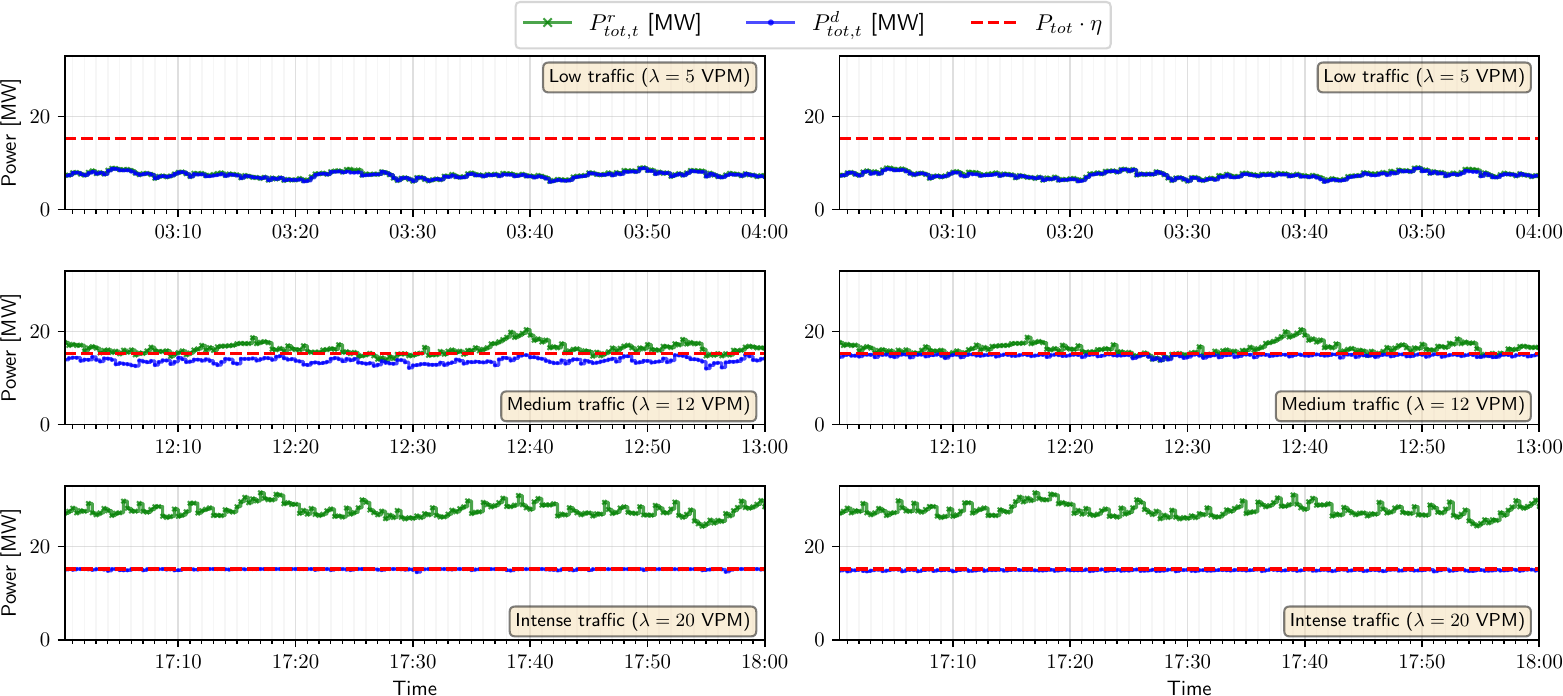}
  \caption{Power allocation comparison: benchmark (left column) vs optimized (right column) for the three Traffic Intensities.
   Green curves represent requested power ($P^{r}_{v,t}$), blue curves show delivered power ($P^{d}_{v,t}$), while the red dashed lines indicate the capacity limit.}
  \label{fig:power_comparison}
\end{figure*}

\Cref{fig:power_comparison} illustrates both the validated behavior of our simulation framework and the optimization opportunities available to the \ac{MPC} controller across three representative traffic scenarios.

In the Low Traffic night scenario (top row), power demand remains well below system capacity.
Both benchmark and optimized configurations successfully satisfy vehicle requests, as evidenced by the near-perfect overlap between requested (green) and delivered (blue) power curves.
This result confirms that when resources are abundant, coordination provides marginal benefit.

The Medium Traffic scenario (middle row) reveals the first optimization opportunity: demand oscillates around the power budget.
Without coordination (left), the benchmark struggles during demand peaks---the delivered power (blue) falls slightly short of the capacity limit, indicating suboptimal resource utilization despite available capacity.
In contrast, the optimized allocation (right) fully utilizes system capacity through dynamic, predictive rebalancing of stripe-level power requests, achieving better matching between supply and demand.

Finally, the High Traffic scenario (bottom row) presents persistent resource scarcity where aggregate demand consistently exceeds system capacity.
Here, both configurations saturate available power, thus resource utilization metrics converge.
In this condition, the analysis of user satisfaction becomes the most interesting one. 

\subsection{Users' Satisfaction Analysis (\socfullfil)}

\begin{figure}[thb]
  \centering
  \subfloat[Without \acsp{VUT}]{%
    \label{fig:energytrajectories_novut}%
    \includegraphics[width=\columnwidth]{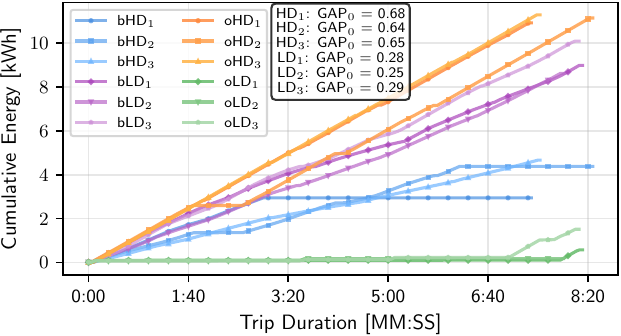}%
  }%
  \hfill
  \subfloat[Vehicles Under Test]{%
    \label{fig:energytrajectories_vut}%
    \includegraphics[width=\columnwidth]{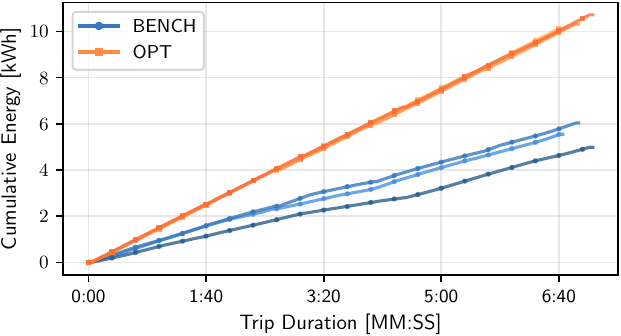}%
  }%
  \caption{Energy trajectories during high-intensity traffic. OPT (orange) consistently prioritizes critical vehicles. 
  (a) Standard scenario, (b) \acsp{VUT} analysis.}
  \label{fig:energytrajectories}
  \vspace{-0.5cm}
\end{figure}

\Cref{fig:energytrajectories} illustrates the temporal evolution of energy accumulated by representative vehicles during high-intensity traffic simulations.
\Cref{fig:energytrajectories_novut} shows the energy allocation over time of 6 selected vehicles with the standard scenario without \acsp{VUT}: the 3 with the highest demand (HD$_{1,2,3}$) and the 3 with the lowest demand (LD$_{1,2,3}$).
The differentiation operated by the optimizer is clear: HD vehicles obtain much more energy at the expense of the LD ones. 
\Cref{fig:energytrajectories_vut} reports the behavior observed on the extremely depleted \acsp{VUT}, and also in this case it is clear that the \ac{MPC} controller is effective in prioritizing \acp{VUT}.

This selective reallocation confirms that, under resource scarcity, the \ac{MPC} framework implements a progressive allocation policy: critical vehicles are prioritized at the expense of well-charged ones, ensuring fairer satisfaction of diverse charging needs.

\begin{figure}[thb]
  \centering
  \subfloat[CDF without \acfp{VUT}]{%
    \label{fig:satisfaction_novut}%
    \includegraphics[width=\columnwidth]{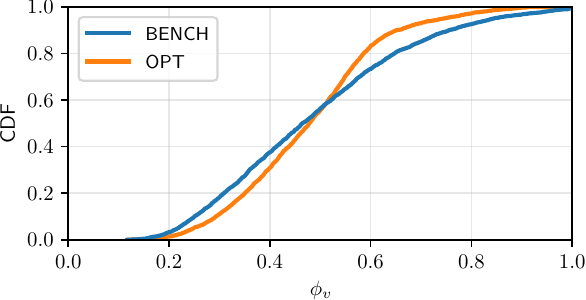}%
  }%
  \hfill
  \subfloat[PDF restricted to \acfp{VUT}]{%
    \label{fig:satisfaction_vut}%
    \includegraphics[width=\columnwidth]{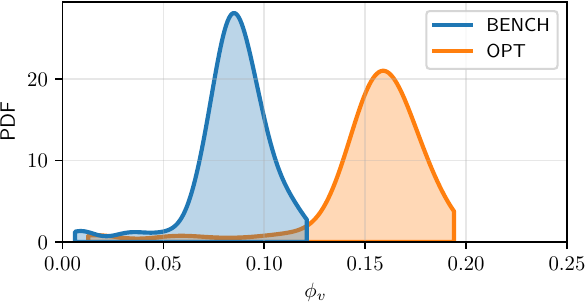}%
  }%
  \caption{CDF (a) and PDF (b) of \ac{SoC} fulfillment (\socfullfil) for high-intensity traffic. 
  OPT (orange) demonstrates progressive allocation: reducing critically unsatisfied vehicles (right tail) at the expense of less-critical ones, particularly pronounced when \acp{VUT} are present.}
  \label{fig:satisfaction_analysis}
  \vspace{-0.5cm}
\end{figure}

\Cref{fig:satisfaction_novut} presents the CDF of the \ac{SoC} fulfillment metric (\socfullfil, defined in~\cref{eq:soc_fulfillment}) for the High Traffic scenario.
The benchmark curve (blue) exhibits remarkably uniform treatment of all vehicles---evidenced by its smooth trajectory, with a long, near-linear middle part. 
The slow convergence to 1.0 indicates that upon exiting the \ac{ITS}, a relatively heavy fraction of vehicles remain critically unsatisfied, i.e., far from their desired \ac{SoC} target.
These vehicles may face range anxiety or the inability to reach home or a conventional charging station without emergency intervention.

In contrast, the optimized allocation (orange) demonstrates progressive, equity-oriented behavior.
Fewer vehicles are highly \textit{un}satisfied, with the orange curve above the blue one for $\socfullfil \geq 0.5$, but also fewer vehicles are highly satisfied (curve inversion at $\socfullfil \approx 0.5$).
This confirms that the \ac{MPC} framework successfully implements a fairness-first policy under resource scarcity, prioritizing the alleviation of critical battery states over maximizing satisfaction of already well-charged vehicles.

\Cref{fig:satisfaction_vut} focuses on the scenario with \acp{VUT}---vehicles entering with nearly depleted batteries ($\socvinit \approx 0$) and maximum charging needs---analyzing only these vehicles. 
The optimized PDF (orange) shows a marked rightward shift compared to the benchmark (blue), indicating that the \ac{MPC} framework successfully prioritizes the most urgent charging needs.
Interestingly, the distributions are nearly completely separated, and it is clear that, given the short time spent in the \ac{DIC} area, full satisfaction cannot be met for these vehicles. 
This gain for critical vehicles is achieved through the strategic reallocation mechanism observed in \cref{fig:satisfaction_novut}: Less-critical vehicles (those with higher initial \ac{SoC} or lower desired targets) absorb less power, redistributing resources to maximize overall satisfaction fairness.
The consistency of this behavior across both scenarios confirms that the urgency-based objective function effectively identifies and prioritizes critical charging needs regardless of explicit vehicle classification.

%

\section{Conclusions and Future Work}
\label{sec:conc}

To the best of our knowledge, this is the first work introducing a system-level analysis of the behavior and performance of future \ac{DIC} systems. 
Results with a non-optimized system show the feasibility of the approach but also highlight that, without an appropriate power management system, the power allocation is suboptimal and can lead to user dissatisfaction and range anxiety. 
This result may be expected as, without management, the system can only split the available power uniformly across all vehicles requesting recharge. 

A power management system requires advanced communication capabilities but, knowing the vehicles' needs, it is possible to optimize power allocation. 
This work proposes and analyzes an optimization framework prioritizing battery-depleted vehicles, using as a target metric the gap between the desired battery level at the exit of the \ac{DIC} area and the actual battery level. 

Results clearly show a significant improvement in user satisfaction (measured by \socfullfil) when the system is overloaded, while in non-overloaded conditions the performance remains unchanged. 
A profitable \ac{DIC} system must hence work most of the time in highly loaded conditions, so that the deployed power is fully utilized. 
Furthermore, market-oriented price differentiation can lead to higher operator revenues, making \ac{DIC} systems economically viable. 

This first analysis and the simulation framework we offer the community open several directions of research parallel to the technological development of \ac{DIC} systems. 
First, different management strategies can be compared as a function of scenario characteristics, including urban vs.\ highway environments, pricing strategies, and stripe placement. 
Second, the actual communication and computational requirements can be analyzed and optimized through the design of a complete communication protocol, considering integration with future \ac{ITS} scenarios, such as \acp{CAM} broadcasting. 
Third, the impact of \ac{DIC} on battery capacity requirements can be investigated. 
If \acp{DIC} become widely adopted, the need for large batteries can be reduced without compromising driving range. 
This in turn reduces vehicle costs and improves the sustainability of \acp{BEV}, potentially triggering a virtuous cycle. 
The framework presented here provides a foundation for exploring these and other research directions in this emerging domain.

\subsection{Open Science}
\label{ss:os}

All code necessary to reproduce the results of this study will be publicly released upon publication at 
\url{https://github.com/lorenzo-ghiro/sumo-wireless-charging}.

%

\section*{Acknowledgments}
\begin{projectacks}
This work was partially supported at the University of Brescia by the European Union and Italian Ministry for Universities and Research National Recovery and Resilience Plan, project ``Sustainable Mobility Center (MOST)'', 2022-2026, CUP D83C22000690001, Spoke N$^o$ 7, ``CCAM, Connected networks and Smart Infrastructures.''
\end{projectacks}

\bibliographystyle{ieeetr}
\bibliography{references, IEEEabrv}

\end{document}